\documentclass[AMS,STIX2COL]{WileyNJD-v2}

\usepackage{graphicx}
\usepackage{float}
\usepackage{amsmath,amsfonts,mathtools}
\usepackage{xcolor}
\usepackage{braket}

\articletype{Conference Proceedings}%

\received{26 April 2016}
\revised{6 June 2016}
\accepted{6 June 2016}

\begin{document}

\title{Ground-State Energy of Uranium Diatomic Quasimolecules with One and Two Electrons}

\author[1]{A. A. Kotov}
\author[1]{D. A. Glazov}
\author[1,2]{A. V. Malyshev}
\author[1]{A. V. Vladimirova}
\author[1]{V. M. Shabaev}
\author[3]{G. Plunien}


\address[1]{\orgdiv{Department of Physics}, \orgname{Saint-Petersburg State University}, \orgaddress{\country{Russia}}}

\address[2]{\orgdiv{NRC <<Kurchatov Institute>>}, \orgaddress{\country{Russia}}}

\address[3]{\orgdiv{Instit\"ut f\"ur Theoretische Physik}, \orgname{Technische Universit\"at Dresden}, \orgaddress{\country{Germany}}}

\abstract[Abstract]{Ground-state energies of the one- and two-electron uranium dimers are calculated for internuclear distances in the range $D=40$--$1000$\,fm and compared with the previous calculations. The generalization of the dual-kinetic-balance approach for axially symmetric systems is employed to solve the two-center Dirac equation without the partial-wave expansion for the potential of two nuclei. The one-electron one-loop QED contributions (self-energy and vacuum polarization) to the ground-state energy are evaluated using the monopole approximation for the two-center potential. Interelectronic interaction of the first and second order is taken into account for the two-electron quasimolecule. Within the QED approach one-photon-exchange contribution is calculated in the two-center potential, while the two-photon-exchange contribution is treated in the monopole approximation.}





\maketitle

\section{Introduction}
Heavy diatomic quasimolecules emerging in ion-ion and ion-atom collisions attracts much interest due to the strong-field phenomena of quantum electrodynamics, such as spontaneous electron-positron pair production \cite{Gerstein69,Pieper69,Zeldovich71,Rafelski78,Greiner85}. While collisions of highly charged ions with neutral atoms have been experimentally observed at the GSI Helmholtz Center for Heavy Ion Research \cite{Verma2006-1,Verma2006-2}, the forthcoming experiments at the GSI/FAIR facilities will enable observation of the heavy ion-ion collisions, up to the encounter of two bare uranium nuclei \cite{FAIR}. 


In this work, we focus on the simple cases of one- and two-electron uranium dimers, U$_2^{183+}$ and U$_2^{182+}$. Starting with the Born-Oppenheimer approximation we consider the ground-state energy of the electron(s) on the basis of the Dirac equation with the Coulomb potential of two nuclei at the fixed internuclear distance $D$. This problem was investigated previously by a number of authors, see, e.g., Refs.~\cite{mueller:73:plb,rafelski:76:plb,rafelski:76:prl,lisin:77:plb,soff:79:pra,lisin:80:plb,deineka:98:os,matveev:00:pan,kullie:01:epjd,Artemyev2010,Tupitsyn2014,Mironova2015,Artemyev2015}. While many of these works relied on the partial-wave expansion of the two-center potential in the center-of-mass coordinate system, the different approaches were also employed. In particular, the application of the Cassini coordinates was studied in Ref.~\cite{Artemyev2010} and the Dirac-Fock-Sturm method was used in Refs.~\cite{Tupitsyn2014,Mironova2015}. The method presented in this work is based on the dual-kinetic-balance approach (DKB) \cite{Shabaev2004} for axially symmetric systems \cite{Rozenbaum2014} with the finite basis set constructed from the B-splines \cite{Johnson88,sapirstein:96:jpb}.


For two-electron system the electron-electron interaction is taken into account within the perturbation theory. The first-order contribution which corresponds to the one-photon-exchange diagram is evaluated as the matrix elements of the interaction operator with the ground-state one-electron wave functions obtained within the dual-kinetic-balance method. The second-order contribution (the two-photon exchange) is evaluated within the rigorous bound-state QED approach, but using the monopole approximation for the binding potential. The one-electron self-energy and vacuum-polarization contributions are also calculated within the monopole approximation.

\section{Method and Results}
In heavy quasimolecules the parameter $\alpha Z \simeq 1$ ($\alpha$ is the fine structure constant and $Z$ is the characteristic nuclear charge) is not small. Therefore, the calculations for these systems should be done to all orders in $\alpha Z$. For this reason, we start with the Dirac equation for the two-center potential ($\hbar=c=m_e=1$), 
\begin{gather}
\label{eq:D}
	\Big[ \vec{\alpha}\cdot\vec{p} + \beta - 1 + V(Z_1,Z_2,\vec{r}) \Big] \Psi_n(\vec{r}) = E_n \Psi_n(\vec{r})
\,,
\\
\label{eq:V}
	V(Z_1,Z_2,\vec{r}) = V_n\big(Z_1,|\vec{r}-\vec{R}_1|\big) + V_n\big(Z_2,|\vec{r}-\vec{R}_2|\big)
\,,
\end{gather}
where $\vec{r}$, $\vec{R}_{1,2}$ are the coordinates of the electron and nuclei, respectively, $Z_{1,2}$ are the nuclear charge numbers, $V_n(Z,r)$ is the nuclear potential. We consider the identical nuclei, $Z_1=Z_2$, with the Fermi model of the nuclear charge distribution.

The solutions of Equation~(\ref{eq:D}) are obtained within the dual-kinetic-balance approach, which allows one to solve the problem of the spurious states. It was developed in Ref.~\cite{Shabaev2004} for the Dirac equation with the central binding potential and later generalized to the axially symmetric case \cite{Rozenbaum2014}. In the latter work, an atom in external homogeneous field was considered. We have adapted this approach for the two-center potential $V(Z_1,Z_2,\vec{r})$ of two nuclei. The considered system is axially symmetric and the $z$-axis is chosen along the internuclear vector $\vec{R_2}-\vec{R_1}$. Therefore, the $z$-projection of the total angular momentum $m_J$ is conserved and the wave function can be written as
\begin{equation}
	\Psi(r,\,\theta,\,\varphi) = \frac{1}{r} \begin{pmatrix}
		G_1(r,\,\theta) e^{ i(m_{J} - \frac{1}{2}) \varphi } \\
		G_2(r,\,\theta) e^{ i(m_{J} + \frac{1}{2}) \varphi } \\
		iF_1(r,\,\theta) e^{ i(m_{J} - \frac{1}{2}) \varphi } \\
		iF_2(r,\,\theta) e^{ i(m_{J} + \frac{1}{2}) \varphi }
    \end{pmatrix}
\,,
\end{equation}
The $(r,\,\theta)$-components of the wave function are represented using the finite-basis-set expansion:
\begin{equation}
	\Phi(r,\,\theta) = \begin{pmatrix}
							G_1(r,\,\theta) \\
							G_2(r,\,\theta) \\
							F_1(r,\,\theta) \\
							F_2(r,\,\theta)
						\end{pmatrix} 
						\cong \sum \limits^{4}_{u = 1} \sum \limits^{N_r}_{i_r = 1} \sum \limits^{N_{\theta}}_{i_{\theta} = 1} C^{u}_{i_r\,i_{\theta}} \Lambda B_{i_r}(r) Q_{i_{\theta}}(\theta) e_u
\end{equation}
where 
\begin{align}
    \Lambda &= \begin{pmatrix}
                1 & -\frac{1}{2}D_{m_J} \\
	            -\frac{1}{2}D_{m_J} & 1
			\end{pmatrix}
\,,\\
  D_{m_J} &= (\sigma_z\cos\theta + \sigma_x\sin\theta)\left(\frac{\partial}{\partial r}-\frac{1}{r}\right) \nonumber
\\
  &+\frac{1}{r}(\sigma_x\cos\theta-\sigma_z\sin\theta)\frac{\partial}{\partial\theta}
\\
  &+\frac{1}{r\sin\theta}\left(im_J\sigma_y+\frac{1}{2}\sigma_x\right), \nonumber
\end{align}
$\big\{B_{i_r}(r)\big\}$ are B-splines, $\big\{Q_{i_{\theta}}\big\}$ are Legendre polynomials of the argument $2\theta/\pi-1$, and $e_u$ are the standard four-component basis vectors, see Ref. \cite{Rozenbaum2014} for details.

\subsection{One-electron system}
First we consider the one-electron U$_2^{183+}$ quasimolecule. The energy of the ground $1\sigma_g$ state is calculated in the range of the internuclear distances $D=40$--$1000$\,fm.
In the calculations we use up to $N_r = 325$ radial basis functions and up to $N_{\theta} = 54$ angular basis functions.
The convergence of the results with respect to the number of basis functions allows us to estimate the numerical uncertainty of the two-center calculations as 50\,eV or less in the entire range of $D$. For comparison, we also consider the monopole approximation, which corresponds to the spherically symmetric term of the multipole expansion of the two-center potential, Equation~(\ref{eq:V}). The calculations in the monopole approximation are performed within the usual DKB approach with by far better accuracy.

\begin{figure}
    \centering
	\includegraphics[width=9cm,height=4.7cm]{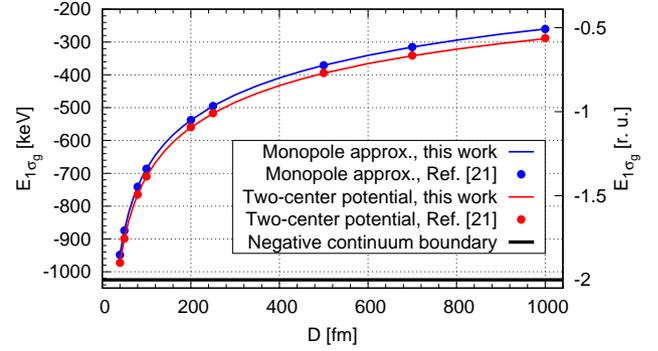}
	\caption{Ground-state binding energy of one-electron U$_2^{183+}$ quasimolecule calculated with the two-center potential (red) in comparison with the monopole approximation (blue). Colored dots (blue and red) correspond to the data from Ref. \cite{Artemyev2015}.}
	\label{fig:gr_st}
\end{figure}
\begin{figure}
    \centering
	\includegraphics[width=9cm,height=4.7cm]{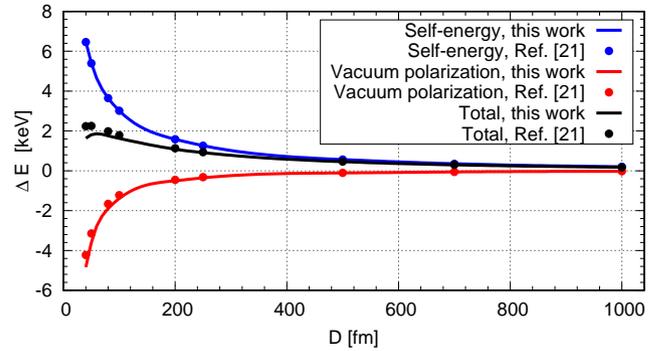}
	\caption{Self-energy (blue) and vacuum-polarization (red) contributions in the monopole approximation. Colored dots (blue and red) correspond to the data from Ref. \cite{Artemyev2015}.}
	\label{fig:se_vp}
\end{figure}

The results are presented in Figure \ref{fig:gr_st} along with the data from Ref.~\cite{Artemyev2015}. One can see that there is the nearly constant gap between the two-center and the monopole results, which varies from $24.5$\,keV at $40$\,fm to $29$\,keV at $1000$\,fm.
We mention also a systematic difference between our values and the values of Ref.~\cite{Artemyev2015}. This difference $E_{1\sigma_g}^{(\text{this work})} - E_{1\sigma_g}^{(\text{ref. \cite{Artemyev2015}})}$ shifts monotonically from $350$~eV at $40$~fm to $-1400$~eV at $1000$~fm.

We also calculate in the monopole approximation the contributions of the self-energy and the vacuum polarization (in the Uehling approximation). To this end, we use the previously developed methods based on the DKB finite basis set, see, e.g., Ref.~\cite{volotka:13:ap} for review and more recent works \cite{artemyev:13:pra,malyshev:14:pra,malyshev:15:pra,malyshev:17:pra}. The results are shown in Figure \ref{fig:se_vp}. The monopole-approximation values from Ref.~\cite{Artemyev2015} are given for comparison. Towards the smaller internuclear distances, one can see the strong enhancement of the QED effects as well as the growing deviation from Ref.~\cite{Artemyev2015}. The reasons of the observed deviations for one-electron system are unclear to us.

\subsection{Two-electron system}
\begin{figure}
	\centering
	\includegraphics[width=9cm,height=4.7cm]{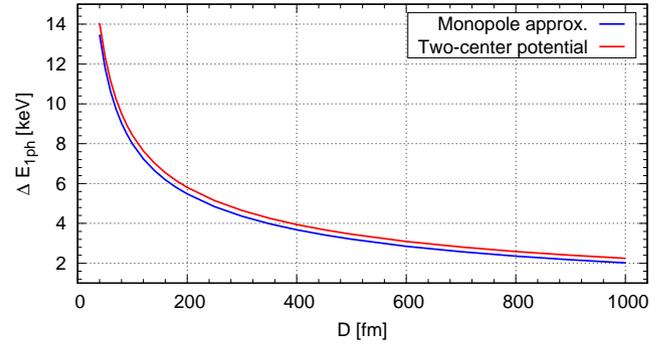}
	\caption{The one-photon-exchange contribution to the ground-state energy of the two-electron U$_2^{182+}$ quasimolecule calculated in the two-center potential and in the monopole approximation of the potential.}
	\label{fig:1ph}
\end{figure}

Let us now consider the two-electron U$_2^{182+}$ quasimolecule. For so high nuclear charge, the independent-electron approximation is a quite reasonable starting point. To zeroth order, the ground-state energy is just the doubled one-electron $1\sigma_g$ energy. The interelectronic interaction can be taken into account by means of the perturbation theory. The first-order contribution corresponds to the one-photon-exchange diagram and is given by
\begin{equation}
\label{eq:dE1ph}
	\Delta E_{1\text{ph}} = \braket{\uparrow\,\downarrow|\,I(0)\,|\uparrow\,\downarrow} - \braket{\uparrow\,\downarrow|\,I(0)\,|\downarrow\,\uparrow}
\,,
\end{equation}
where the arrows denote the $1\sigma_g$ states with $m_J=\pm1/2$. We use the interelectronic-interaction operator in the Feynman gauge,
\begin{equation}
	I(\omega) = \alpha \frac{1-\vec{\alpha}_1\cdot\vec{\alpha}_2}{r_{12}} \exp\big(i\,|\omega|\,r_{12}\big)
\,,
\end{equation}
where $\omega$ is the energy of the exchanged photon and $\vec{\alpha}_i$ is the vector of the Dirac $\alpha$-matrices, which act on the $i$th electron. We note that the evaluation of the matrix elements in Equation~(\ref{eq:dE1ph}) is rather involved in the present case. In contrast to the spherically symmetric systems, the partial-wave expansion of $I(\omega)$ yields nonzero contributions up to the infinite angular momenta and has to be truncated according to its convergence. In addition, the numerical integration over the four variables, $r_1$, $r_2$, $\theta_1$, and $\theta_2$ is implied. The results for the one-photon-exchange contribution are displayed in Figure~\ref{fig:1ph}.

\begin{figure}
	\centering
	\includegraphics[width=9cm,height=4.7cm]{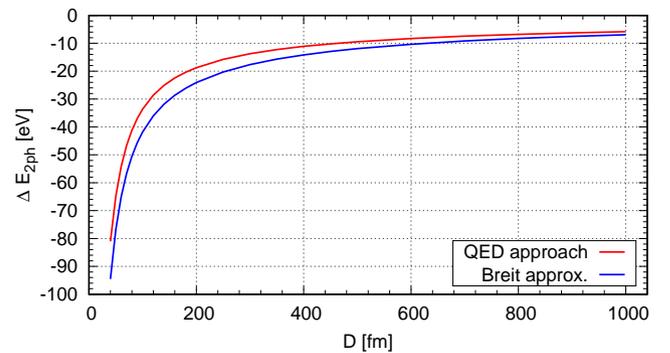}
	\caption{The two-photon-exchange contribution to the ground-state energy of the two-electron U$_2^{182+}$ quasimolecule calculated in the monopole approximation calculated within the rigorous QED approach and within the Breit approximation.}
	\label{fig:2ph}
\end{figure}

The second-order (the two-photon-exchange) contribution is calculated within the monopole approximation. The calculations are performed within the rigorous QED approach valid to all orders in $\alpha Z$ and within the Breit approximation. We refer again to the review \cite{volotka:13:ap} and to the more recent papers \cite{artemyev:13:pra,malyshev:14:pra,malyshev:15:pra,malyshev:17:pra} for details on the general formulae and the computational methods for this problem. The results obtained for the ground state of the U$_2^{182+}$ quasimolecule are presented in Figure~\ref{fig:2ph}.


\section{Conclusion}
In this work, the application of the dual-kinetic-balance approach to the two-center Dirac equation with axial symmetry has been demonstrated. The ground-state energy of one-electron uranium-uranium quasimolecule has been calculated and compared with the the monopole-approximation results and to the previous calculations \cite{Artemyev2015}. The one-loop QED corrections have been obtained within the monopole approximation and compared also with Ref.~\cite{Artemyev2015}.

For the ground state of two-electron uranium-uranium quasimolecule, the one-photon-exchange contribution has been calculated for the two-center potential. The two-photon-exchange has been calculated within the monopole approximation for the binding potential. A significant increase of these contributions at small internuclear distances has been observed. 

The calculations presented in this work and their extension to the excited states and different nuclei are relevant for interpretation of the quasimolecular radiation spectra in heavy-ion collisions.


\section*{Acknowledgements}

Valuable discussions with Ilia Maltsev, Leonid Skripnikov, and Ilya Tupitsyn are gratefully acknowledged.
The work was supported by RFBR (Grants No. 16-02-00334 and No. 19-02-00974), by SPbSU-DFG (Grant No.~11.65.41.2017 and No.~STO 346/5-1), by SPbSU (COLLAB 2018: 34824940), by the FAIR-Russia Research Center, and by TU Dresden (DAAD Programm Ostpartnerschaften). A. V. Malyshev and V. M. Shabaev acknowledge the support from the Foundation for the advancement of theoretical physics and mathematics "BASIS".

\end{document}